IMMUNOLOCALIZATION OF $NA^+,K^+$-ATPASE IN THE BRANCHIAL CAVITY DURING THE EARLY DEVELOPMENT OF THE CRAYFISH <u>ASTACUS LEPTODACTYLUS</u> (CRUSTACEA, DECAPODA).

Jean-Hervé Lignot [1] *, Gregorius Nugroho Susanto [2], Mireille Charmantier-Daures [3], Guy Charmantier [3]

(1) Centre d'Ecologie et de Physiologie Energétiques CEPE-CNRS, UPR 3090

23 Rue Becquerel,

67087 Strasbourg, France.

Tel: (333) 8810 6938       Fax: (333) 8810 6906       Email: J-H.Lignot@c-strasbourg.fr

(2) Department of Biology,

Faculty of Mathematics and Natural Sciences,

University of Lampung, Jl. Sumantri Brojonegoro 1,

Bandar Lampung, Indonesia, 35145.

(3) Adaptation Écophysiologie au cours de l'Ontogénèse

UMR 5000, GPIA

Université Montpellier II

CP 092, Place Eugène Bataillon,

F-34095 Montpellier Cedex 05, France.

Tel.: (334) 6714 3672       Fax: (334) 6714 9390       E-mail: charmantier@univ-montp2.fr

*: corresponding author

ABSTRACT

The ontogeny of osmoregulation was examined in the branchial cavity of embryonic and early postembryonic stages of the crayfish Astacus leptodactylus maintained in freshwater, at the sub-cellular level through the detection of the $Na^+,K^+$-ATPase. The embryonic rate of development was calculated according to the eye index (EI) which was 430-450 µm at hatching. The distribution of the enzyme was identified by immunofluorescence microscopy using a monoclonal antibody IgGα5 raised against the avian α-subunit of the $Na^+,K^+$-ATPase.

Immunoreactivity staining indicating the presence of $Na^+,K^+$-ATPase appeared in the gills of late embryos (EI ≥ 400 µm), i.e. a few days before hatching time, and steadily increased throughout the late embryonic and early postembryonic development. The appearance of the enzyme correlates with the ability to osmoregulate which also occurs late in the embryonic development at EI 410-420 µm and with tissue differentiation within the gill filaments. These observations indicate that the physiological shift from osmoconforming embryos to hyper-regulating late embryos and post-hatching stages in freshwater must originate partly from the differentiation in the gill epithelia of ionocytes which are the site of ion pumping, as suggested by the location of $Na^+,K^+$-ATPase. Only the gills were immunostained and a lack of specific staining was noted in the lamina and the branchiostegites. Therefore, osmoregulation through $Na^+$ active uptake is likely achieved in embryos at the gill level; all the newly-formed gills in embryos function in ion regulation; other parts of the branchial chamber such as the branchiostegites and lamina do not appear to be involved in osmoregulation.

INTRODUCTION

Crayfish are decapod crustaceans fully adapted to freshwater, where they spend their entire life-span including, reproduction and development. Their ion regulation is therefore challenged by low medium osmolality and by low external concentrations of sodium, chloride and calcium. The concentrations of ions in crayfish fluids and tissues and hemolymph osmolality are maintained within set ranges. For instance, hemolymph osmolality is approximately 415-420 mosm.kg$^{-1}$ in the adults of <u>Astacus leptodactylus</u> kept in freshwater (Bielawski, 1964; Holdich et al., 1997; Susanto and Charmantier, 2000). Crayfish are thus strong hyper-osmoregulators in freshwater. This is achieved through three main mechanisms, including low tegument water and ion permeability, an active ion uptake from food and via specialised branchial chamber ion-transporting tissues, and the production of hypotonic urine. Due to the very low ionic concentrations in freshwater, an efficient transport mechanism is required in order to achieve the net uptake of ions from the medium.

In marine, estuarine and freshwater crustaceans, inward pumping of sodium ions out of specialised cells (ionocytes) into the haemolymph is driven by the activity of Na$^+$,K$^+$-ATPase, a key enzyme in ion transport which uses ATP as the source of energy (De Renzis and Bornancin, 1984). High levels of activity of this enzyme are evident in osmoregulatory structures located in gills (see reviews in Mantel and Farmer, 1983; Gilles and Péqueux, 1985; Péqueux and Gilles, 1988; Lucu, 1990; Taylor and Taylor, 1992; Péqueux, 1995; Freire and McNamara, 1995; Zare and Greenaway, 1998; Towle and Weihrauch, 2001), but also in extrabranchial organs of some seawater and freshwater crustaceans. Among these organs are the pleurites (Talbot et al., 1972; Felder et al., 1986; Bouaricha et al., 1994), the branchiostegites (Talbot et al., 1972; Felder et al., 1986; Bouaricha et al., 1994; Haond et al., 1998; Lignot et al., 1999; Lignot and Charmantier, 2001), the epipodites (Kikuchi and Matsumasa, 1993; Bouaricha et al., 1994; Dunel-Erb et al., 1997; Haond et al., 1998; Lignot et al., 1999; Lignot and Charmantier, 2001), all located in the branchial chambers, and a few other organs situated outside these cavities in non-decapod crustaceans (Hootman et al., 1972; Conte et al., 1972; Holliday et al., 1990; Aladin and Potts, 1995; Kikuchi and Matsumasa, 1995, 1997; Hosfeld and Schminke, 1997).

In freshwater crayfish, the site of ion uptake was firstly identified in the ionocytes of the crayfish <u>Astacus pallipes</u> (Fisher, 1972), these cells being uniformely distributed among the gill pairs (Dickson and Dillaman, 1985). A spatial separation between gas-exchange and ion-transport functions was later determined within each gill, individual gill filaments being involved either in respiration (with thin epithelia) or ion transport (with thicker epithelia and numerous ionocytes, showing characteristic ion-transporting features such as apical microvilli, basolateral infoldings associated with numerous mitochondriae) (Dickson et al., 1991; Dunel-Erb et al., 1997). $Na^+,K^+$-ATPase activity was also higher in the transporting filaments than in those designated as respiratory (Dickson et al., 1991). Additionally, ion-transporting cells have been described in the epipodites (denominated lamina in crayfish) but they differ from those of the gill filaments in having membrane infolding systems on the apical side of the cells only. This suggested they had a different role in osmoregulation, perhaps in anion transport (Dunel-Erb et al., 1997; Barradas et al., 1999a, b).

Despite detailed morpho-functional studies on the gill filaments of adult or large juvenile freshwater crayfish, no data on the distribution and characterisation of ion-transporting tissues are available yet in embryos and young juveniles (Dunel-Erb et al., 1982; Wheatly and Gannon, 1995; Susanto and Charmantier, 2000, 2001). The ontogeny of osmoregulation however, has been clearly demonstrated as a major adaptive process to environmental salinity, the establishment of a species in a given habitat, depending on the ability of each of its developing stages to adapt to the environment. Furthermore, studies on the ontogeny of osmoregulation in crustaceans are mostly dedicated to marine and estuarine decapod species (reviews in Charmantier, 1998; Charmantier and Charmantier-Daures, 2001).

In the freshwater crayfish A. leptodactylus, embryos develop in eggs which are attached to the pleopods of the female and are thus directly exposed to the external medium. The direct development of crayfish leads to a larval development in the egg and juveniles at hatching (i.e., no free-living larval stages) (Zehnder, 1934; Celada et al., 1985, 1987). The acquisition of the ability to hyper-osmoregulate occurs late in the embryonic development, only hours before hatching. First stage juveniles are therefore able to osmoregulate and their ability increases during the postembryonic development (Susanto and Charmantier, 2000, 2001). Establishing the tissue and cellular differentiation towards ion transport during the embryonic and early post-embryonic development is of special interest with respect to further functional and morphological identification of ion transport.

The general aim of the present study is to investigate the ontogeny of the osmoregulatory structures and their functionality during the late embryonic and the early post-embryonic development of the freshwater crayfish, A. leptodactylus Escholtz, 1823. This study follows the physiological ontogenetic study of osmoregulation in the same species (Susanto and Charmantier, 2000, 2001) and is aimed at further ascertaining the ontogenetic involvement in osmoregulation of ion-transporting epithelia within the branchial cavities. The observations were conducted in embryos, juvenile stages I to III and in one-year old juveniles through histology and, mainly, by immunofluorescence on-section labelling using a monoclonal antibody IgG$\alpha$5 raised against the avian $\alpha$-subunit of the Na$^+$,K$^+$-ATPase.

MATERIALS AND METHODS

Animals

Adult berried female A. leptodactylus imported from Turkey and obtained from a commercial retailer in Saint Guilhem-Le-Desert, Hérault, France were transported to the Montpellier laboratory and maintained in 40-L individual compartments containing aerated dechlorinated tap water (maintained at 19±0.5°C and half-changed daily). The composition (in mmol.L$^{-1}$) of this water was Na$^+$ (0.12), K$^+$ (0.04), Ca$^{++}$ (5.70), Mg$^{++}$ (0.29), Cl$^-$ (0.08), NO3$^-$ (0.06) and SO4$^{--}$ (0.61); its osmolality was 11 mosm.kg$^{-1}$. A 12 h light: 12 h dark photoperiod was maintained. Berried females were fed three times a week with thawed mussels.

Embryonic development was determined according to Zehnder's (1934) description. Following the eyes' appearance, the eye index (EI) was calculated according to the method of Perkins (1972) adapted to the crayfish (EI = half sum of length and width of the pigmented part of the eye, in µm). The value of EI was 430-450 µm immediately before hatching. For the subsequent experiments, embryos were randomly chosen from different batches and were grouped into five categories (EI = 300; 360; 400; 411; 420 µm), determined according to specific events in the ontogeny of osmoregulation (Susanto and Charmantier, 2001).

Following hatching, the lecitotrophic juveniles in stage I were mass-reared with each female in the 40-L tanks. Juvenile stages II and III were maintained in individual compartments provided with aerated, dechlorinated and recirculated (Eheim systems) tap water under the same conditions as above. Frozen Artemia sp. were distributed as food daily. Following stage III, chopped flesh mussels replaced Artemia for feeding the young crayfish. The average durations of successive juvenile stages were 4-5 d (stage I), 7-8 d (stage II), 11-13 d (stage III). The sampled juveniles were staged according to morphological criteria (Payen, 1973) and to dates of their molts, which were recorded. For all the studied juvenile stages, observations were conducted on intermolt stage C individuals (Drach and Tchernigovtzeff, 1967). For each category (embryos with EI between 300 and 420 µm, juvenile in stages I to III, and one-year old juveniles), five to ten individuals were randomly chosen for the structural and immunocytochemical study.

Histology

Embryos extracted from the eggs under a dissecting microspcope were fixed for 24 hrs in Bouin's fixative. Specimens were fully dehydrated and embedded in paraffin. Sagittal sections (3 µm) cut on a Leitz Wetzlar microtome were collected on glass slides and were stained with picro-indigo-carmine stain (Martoja and Martoja, 1967).

Scanning electron microscopy

Embryos were critical-point dried with liquid $CO_2$ in a Baltec CPD 030 critical-point dryer and dehydrated. After being mounted on stubs, the samples were coated with silver in a Baltec SOD 050 ion coater, and examined with a Jeol JSM-6300f scanning electron microscope operated at 15 kV.

Immunofluorescence light microscopy

Embryos and cephalothoraces of juvenile stages I to III were fixed and processed as for the structural study (see above), but with sagittal sections (3 µm) collected on poly-L-lysine-coated slides. The technique for the immunocytochemical demonstration of $Na^+,K^+$-ATPase in epithelial cells was similar to the procedures of Ziegler (1997), Lignot et al. (1999) and Lignot and Charmantier (2001). Sections were preincubated for 10 min in 0.01 mmol Tween 20, 150 mmol NaCl in 10 mmol phosphate buffer, pH 7.3, then treated with 50 mmol $NH_4Cl$ in phosphate-buffered saline (PBS), pH 7.3, for 5 min to mask free aldehyde groups of the fixative. The sections were washed in PBS and incubated for 10 min with a blocking solution (BS) containing 1 % bovine serum albumin (BSA) and 0.1 % gelatin in PBS. Droplets (10 µl) of the primary antibody (monoclonal antibody IgGα5, raised against the avian α-subunit of the Na$^+$,K$^+$-ATPase developed and given by Dr. Fambrough) diluted in PBS at 20 µg.mL$^{-1}$ were placed on the sections and incubated for 2 h at room temperature in a wet chamber. Control sections were incubated in BS without the primary antibody. After washing six times for 5 min in BS to remove unbound antibodies, the sections were incubated for 1 h in droplets of secondary antibody [FITC-conjugated goat anti-mouse IgG (H&L) (Jackson Immunoresearch, West Baltimore, MD)]. Following extensive washing in BS (six times for 5 min), sections were mounted in 80 % glycerine, 20 % PBS plus 2 % N-propyl-gallate to retard photobleaching. Sections were examined with a fluorescence microscope (Leitz Diaplan coupled to a Ploemopak 1-Lambda lamp) equipped with the appropriate filter set (450-490 nm band-pass excitation filter) and a phase-contrast device.

RESULTS

General morphology and histological structure

Each of the two branchial cavities located on both sides of the cephalothorax contains 20 trichobranchiate gills and 7 convoluted lamina. Each gill connected to the pleurite possesses short filaments attached to a central axis (Fig. 1), and lamina that are associated with the podobranchs (except on the first maxilliped) are lamellar V-shaped organs (Fig. 1).

These organs are already present in late embryos and show a progressive differentiation from embryos to juvenile stages I-III. Gills appear in embryos with EI around 300 µm and they contain masses of undifferentiated cells (Fig. 2A). In embryos with EI ≤ 400 µm, the branchial cells develop progressively into regular branchial epithelium (Fig. 2A, B, C) consisting of a single layer of cells, 20 to 30 µm thick (Fig. 2C). In embryos with EI ≥ 400 µm and in juvenile stage I, afferent and efferent vessels become separated by a thin septum (Fig. 2D, E). The lamina is already present in embryos before hatching and it is limited by a thin and regular epithelium covering both sides of the organ (Fig. 2A, B, C). A central and elongated hemolymph lacuna also appears in late embryonic stages and separates the two epithelial layers (Fig. 2C). In all the studied embryonic stages, the branchiostegites possess an outer limiting epithelium containing numerous nuclei along the outer-cuticle, a vascular space and a loose inner-side epithelium containing few nuclei and bordered by a thin cuticle on the side limiting the branchial cavity (Fig. 2A, B, C).

Immunofluorescence light microscopy

The Bouin fixation and paraffin embedding procedures yielded good antigenicity as observed with the fluorescent micrographs (Fig. 3A, C, E, G and 4A, C, E, G) and good structural preservation as observed with phase-contrast pictures (Fig. 3B, D, F, H and 4B, D, F, H). Immunofluorescence microscopy also showed consistent results within embryonic and post-embryonic stages. Controls showed no specific binding within gill, branchiostegite and epipodite epithelia (not illustrated).

In embryos at EI ≈ 300 μm, the branchiostegite and lamina showed no specific staining. In gills already present at that stage, immunoreactivity was very weak with patchy staining in only some gill filaments (Fig. 3A). Similar observations were made in embryos at EI = 360 μm. At EI = 400 μm, heavier fluorescent staining indicating the presence of the $Na^+$, $K^+$-ATPase was observed in all the gills and in most of the gills filaments (Fig. 3C). The branchiostegite and epipodites showed no staining, as for all the following late embryonic and postembryonic stages. At EI = 411 μm, copious staining was observed in the gill filaments but also along the stem of the gills (Fig. 3E). At EI = 420 μm, i.e. a few hours before hatching, the fluorescence was then restricted to the thickest epithelia lining only some of the gill filaments (Fig. 3G).

After hatching, in early juveniles, the $Na^+$, $K^+$-ATPase was observed only in the ion-transporting type branchial filaments (Fig. 4A to 4G), which were characterized by thick striated epithelia. Immunoreactivity also appeared restricted to gill filaments located at the tip of the gills (Fig. 4C) and heavier staining could be observed with the progress of the crayfish development (Fig. 4A to 4G). In all these stages presenting specific fluorescence, the $Na^+,K^+$-ATPase was mainly located along the basal lamina of the thick epithelial layers as illustrated in stages I and III and in one-year old juveniles (Fig. 4A to 4G).

DISCUSSION

Sodium-potassium adenosinetriphosphatase ($Na^+,K^+$-ATPase), the major driving force for the active transport of electrolytes in various tissues, consists of a catalytic $\alpha$-subunit of approximately 100 kDa and a smaller less functionally characterized $\beta$-subunit (Horisberger et al., 1991). The mouse monoclonal IgG$\alpha$5, raised against the avian $\alpha$-subunit of the $Na^+,K^+$-ATPase, recognizes three $\alpha$ isoforms ($\alpha$1, $\alpha$2, $\alpha$3) and cross-reacts with the $\alpha$-subunits of the $Na^+,K^+$-ATPase of several fish and insect epithelia (Lebowitz et al., 1989; Baumann and Takeyasu, 1993; Baumann et al., 1994; Just and Waltz, 1994; Witters et al., 1996). In crustaceans, this antibody specifically recognizes a single band of about 90-110 kD, as observed in the terrestrial isopod Porcellio scaber (Ziegler, 1997) and in the European decapod lobster Homarus gammarus (Lignot et al., 1999; Lignot and Charmantier, 2001). In the present immunocytochemical study, the mouse monoclonal IgG$\alpha$5 appears to recognizes its epitope in the branchial cavity of the developing freshwater crayfish, Astacus leptodactylus. This high antibody specificity may be attributable to the conservation of the $\alpha$-subunit of the protein in the animal kingdom during the course of evolution, as illustrated in Artemia franciscana by the analysis of its amino acid sequence, which is about 80 % similar to the vertebrate sequence (Macias et al., 1991), whereas there is approximately 90 % similarity between birds, fish and mammals (Schull et al., 1985; Kawakami et al., 1985; Takeyasu et al., 1988).

In freshwater crayfish, gills appear in embryos with EI around 300 μm and develop progressively with regular branchial epithelia and afferent and efferent vessels presumably fully functional prior hatching. Within the cell layer of the epithelium, few basolateral infoldings, but numerous vesicles sometimes associated with mitochondria, can be observed (Susanto, 2000). Well-characterized ionocytes (specialized cells with apical microvilli and basolateral infoldings associated with numerous mitochondria) were mostly observed after hatching in juvenile stage I but without the differentiation between respiratory and ion-transporting filaments within each gill (Susanto, 2000) which has been reported in adult crayfish such as Procambarus clarkii (Dickson et al., 1991) and Astacus leptodactylus (Dunel Erb et al., 1997; Barradas et al., 1999a, b). In the crayfish late embryos and post-hatch juvenile stages, however, we report here that $Na^+,K^+$-ATPase was found along the branchial filaments, mostly at the tips of the gills and at the basal side of the epithelial cells. This observation indicates that the $Na^+,K^+$-ATPase appears at an early stage in the crayfish development, well before the morpho-functional differentiation between branchial ion-transporting and respiratory epithelia. Only one type of epithelial cell thus might be involved in both functions, as already observed in some freshwater or terrestrial species (Taylor and Greenaway, 1979; Farelli and Greenaway, 1992). According to another hypothesis, the diffusive gas exchanges might be effected through the entire body surface of the embryos until the differentiation of respiratory gill filaments. The presence of the enzyme in the gills of decapod crustaceans has been documented in numerous freshwater and euryhaline species (reviews in Mantel and Farmer, 1983; Gilles and Péqueux, 1985; Péqueux and Gilles, 1988; Lucu, 1990; Taylor and Taylor, 1992; Péqueux, 1995; Freire and McNamara, 1995; Towle and Weihrauch, 2001). It is mainly located in the posterior gills of Brachyurans, the anterior gills being mainly respiratory (Towle and Kays, 1986; Péqueux et al., 1988 in Taylor and Taylor, 1992; Pierrot, et al., 1995) and in one species of isopod (Holliday et al., 1988). In some crustaceans, this partitioning between gas-exchange and ion-transport functions has also been demonstrated within each gill as in the euryhaline crabs Callinectes sapidus, Carcinus maenas (Neufeld et al., 1980; Towle and Kays, 1986), Eriocheir sinensis (Péqueux et al., 1988 in Taylor and Taylor, 1992) and Pachygrapsus marmoratus (C. Spanings-Pierrot personal communication). Ionocytes have also been described in the lamina of adult A. leptodactylus (Dunel-Erb et al., 1997; Barradas et al., 1999a, b) but they lack $Na^+,K^+$-ATPase. They differ from those found in the gill filaments of the crayfish or in the lobster epipodite and branchiostegite epithelia (Haond et al., 1998; Lignot et al., 1999), in having membrane infolding systems with an opposite orientation (i.e., on the apical side of the cells), suggesting therefore a different role in osmoregulation. Moreover, differences in ionic permeabilities of the cuticle in A. leptodactylus have also been demonstrated depending on whether the lamina or the gill filaments are considered. As hypothesized by Dunel-Erb et al. (1997), the lamina with a cuticle selectively permeable to $Cl^-$ and $OH^-$ (Avenet and Lignon, 1985) and whithout $Na^+,K^+$-ATPase (Barradas et al., 1999b and this study) might be involved in the movements of anions. The ion-transporting filaments which possess $Na^+,K^+$-ATPase (Barradas et al., 1999b and this study) and possibly $H^+$ V-ATPase (Zare and Greenaway, 1997, 1998) might be involved in the movements of $Na^+$, $K^+$, $NH_4^+$, and $Ca^{2+}$ to which the cuticle is highly permeable (Avenet and Lignon, 1985; Lignon and lenoir, 1985; Lignon and Péqueux, 1990).

The appearance of $Na^+,K^+$-ATPase in the gill ionocytes of late embryos of A. leptodactylus starting at EI 400 µm correlates well with the available knowledge on the ontogeny of osmoregulation in embryos of this species (Susanto and Charmantier, 2001). The acellular egg envelopes of A. leptodactylus apparently do not include sites for active ion transport and may be impermeable, limiting therefore water loss and ion invasion (Susanto and Charmantier, 2001). In experiments conducted by these authors, the egg membranes were cut open and the embryos were directly exposed to freshwater. Hemolymph osmolality was then measured at regular time intervals. For EI ranging from 300 to 390 µm, the embryos' hemolymph osmolality dropped sharply from 360-380 to 100-150 mosm.kg$^{-1}$ within 1-2 h. The embryos swelled, outer limiting teguments burst, resulting in 100 % mortality due to a sudden massive water influx. At EI 411 µm, hemolymph osmolality decreased more slowly and stabilized for 5 h at 220-250 mosm.kg$^{-1}$, then decreased further before the death of embryos. At EI 425 µm, the hemolymph osmolality stabilized within 1 h at 280-290 mosm.kg$^{-1}$ (a value found in naturally-hatched juveniles) and the young crayfish survived and developed into juvenile stages. Thus, the ability to osmoregulate appears late in the embryonic development of the crayfish and correlates well with the occurrence of ionocytes on the gills, in which Na$^+$, K$^+$-ATPase mediates active ion uptake. We report here that the enzyme is expressed in embryos at EI 400-420 µm, i.e. at the time when they become fully able to hyper-osmoregulate in freshwater. Later in development, the Na$^+$, K$^+$-ATPase is expressed or an increasing number of the gill filaments of juveniles and the capacity to hyper-osmoregulate increases concomitantly (Susanto and Charmantier, 2000).

In conclusion, in the freshwater crayfish A. leptodactylus, the embryos are unable to osmoregulate during most of their developement. They are osmotically protected from freshwater by the egg envelopes according to a mechanism which still has to be fully explained (Susanto and Charmantier, 2001). A cellular differentiation of the gills and lamina and the appearance of Na$^+$,K$^+$-ATPase in the gill filaments occur in late embryos starting at EI 400 µm, i.e. a few hours or days (according to temperature) before hatching. We consider that this molecular and cellular differentiation is one of the main adaptations leading to the occurrence of the ability to hyper-osmoregulate in freshwater, which appears in late embryos and that results in the ability of newly-hatched juveniles to live under the osmotic stress of freshwater. Young juveniles emerging from the eggs are then able to hyper-osmoregulate, and their ability increases with a further enhancement of $Na^+,K^+$-ATPase activity in gill filaments through postembryonnic development.

ACKNOWLEDGEMENTS

We wish to thank Dr D.M. Fambrough for his generous gift of the anti- $Na^+,K^+$-ATPase antibody, Ms E. Grousset and V. Richard for their technical help.

Fig. 1. <u>Astacus leptodactylus</u>, juvenile I. Right branchial chamber of a newly-hatched crayfish. The branchiostegite (bst) has been cut to reveal the six podobranchs and the epipodites connected to them. The arthrobranchs are not visible as they lie beneath the podobranchs. bst, branchiostegite; gf: gill filament, lm: lamina, sgt: scaphognatite. Bar: 100 µm.

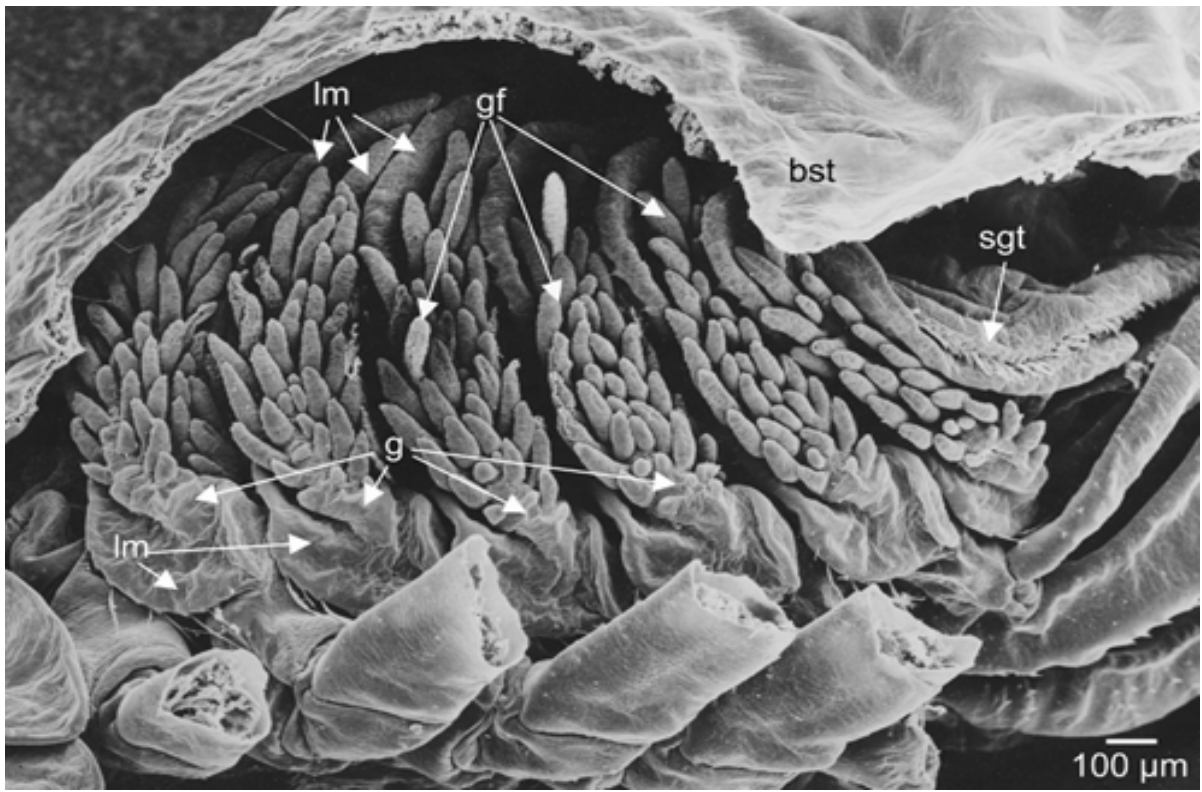

Fig. 2. <u>Astacus leptodactylus</u>. Embryos (A: EI = 367 μm; B: EI = 400 μm; C: EI = 420 μm) and juvenile I (D). Transverse sections of the branchial cavity. af, afferent vessel; bc, bordering cell; bst, branchiostegite; cu, cuticle; ef, efferent vessel; gf, gill filament; lm, lamina; n, nucleus; pd, podocyte, s: septum. Bars: 100 μm.

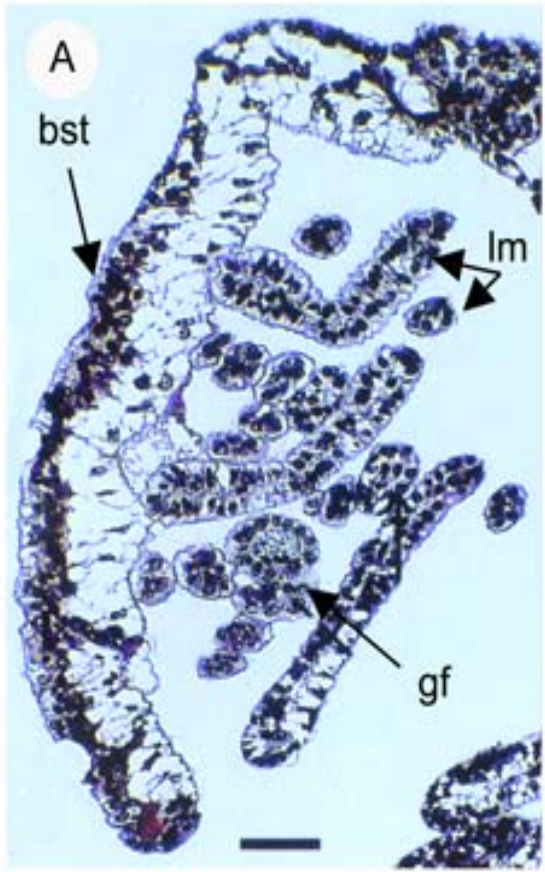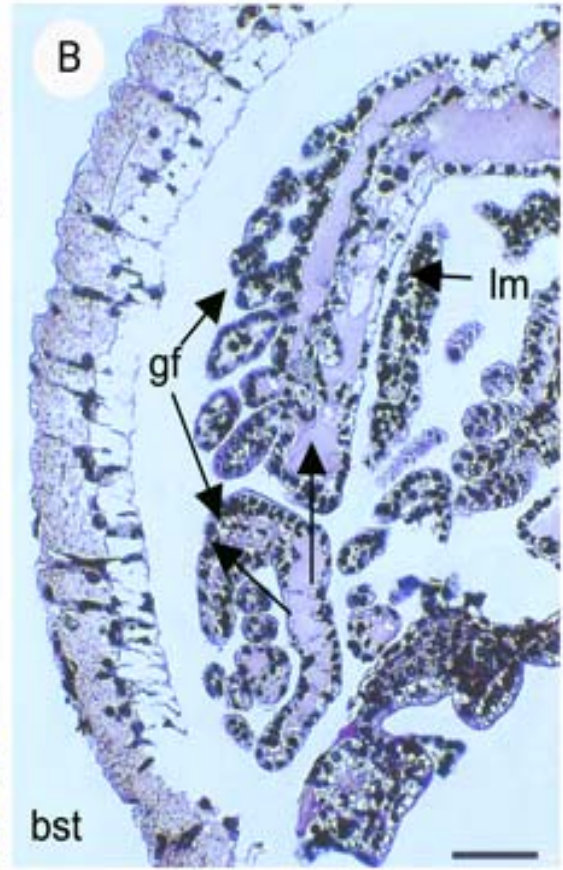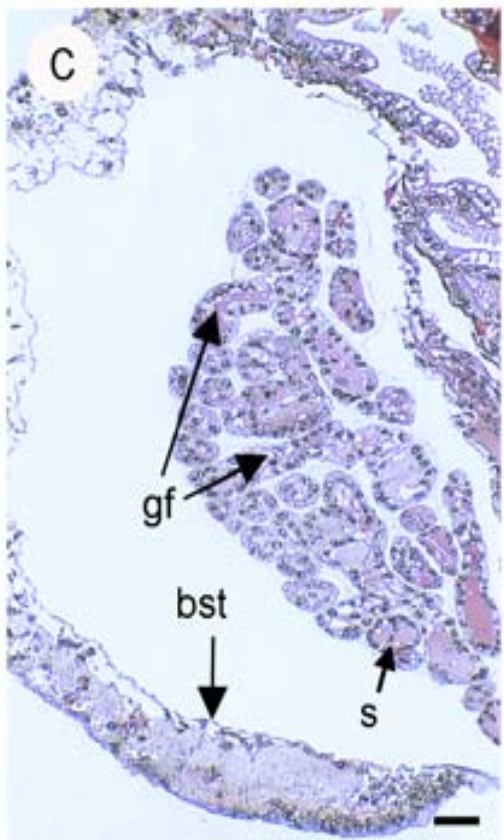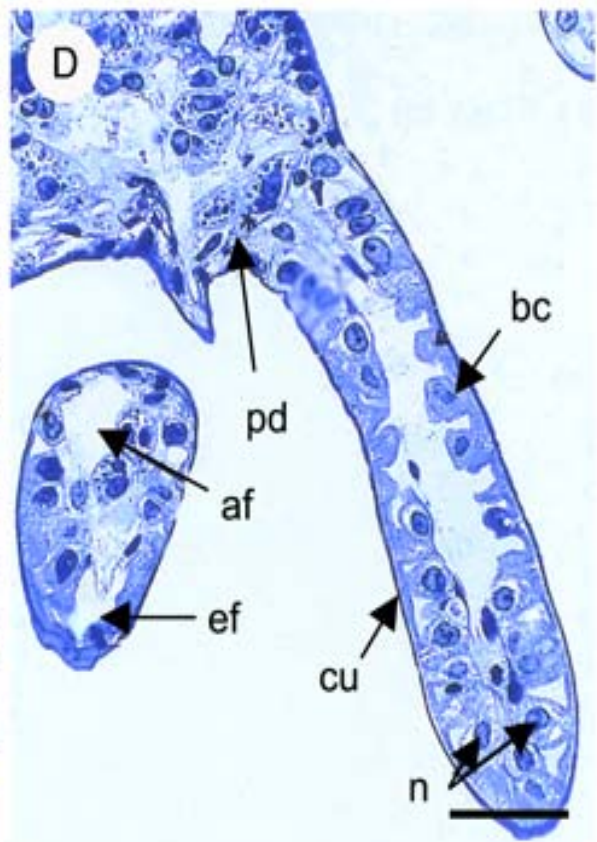

Fig. 3. <u>Astacus leptodactylus</u>, embryos (A, B: EI = 300 μm; C, D: EI = 400 μm; E, F: EI = 411 μm; G, H: EI = 420 μm). Immunolocalization of the $Na^+,K^+$-ATPase on thin sections of the branchial chamber. (A, C, E, G): fluorescent micrographs; (B, D, F, H): corresponding phase-contrast pictures. bst, branchiostegite; cu, cuticle; gf, gill filaments; hl, hemolymph lacuna; iep, inner-side limiting epithelium; lm, lamina; n, nucleus; oep, outer-side limiting epithelium; pc, pillar cells. On the fluorescent micrographs, arrows indicate Na-K-ATPase reactivity. Bars: 40 μm.

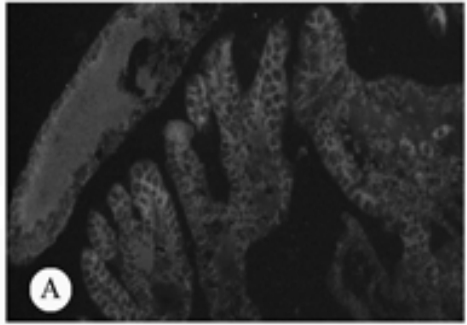 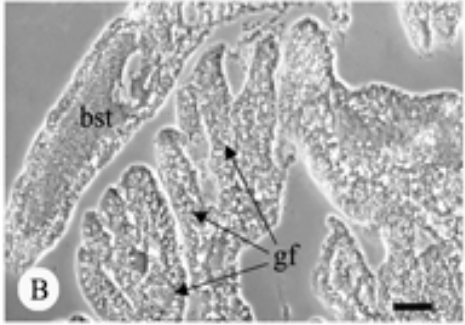
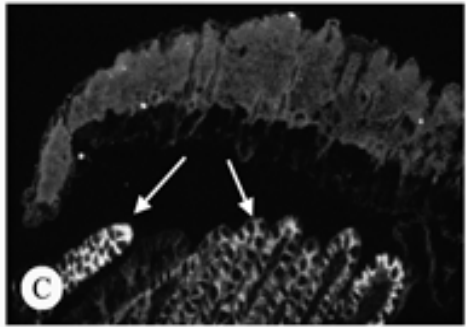 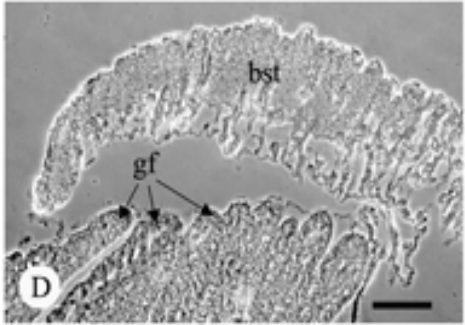
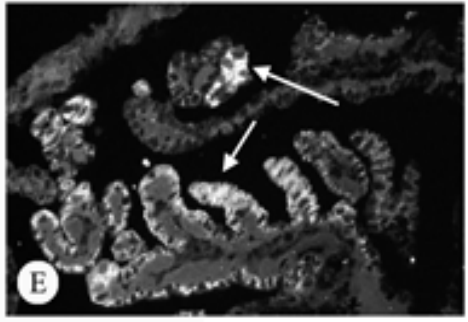 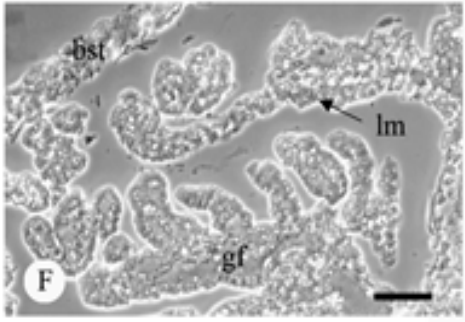
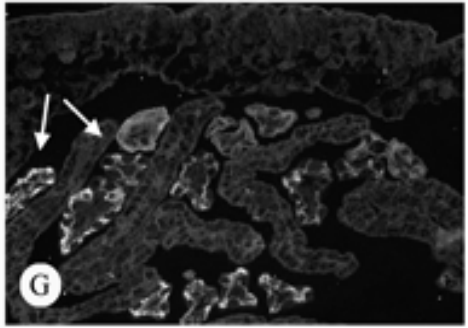 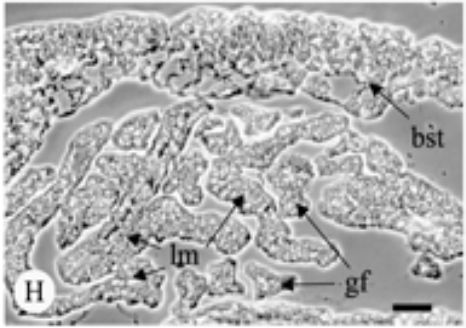

Fig. 4. <u>Astacus leptodactylus</u>, juveniles. Immunolocalization of the Na$^+$,K$^+$-ATPase on thin sections of the branchial chamber. (A, B, C, D): juvenile I; (E, F): juvenile III; (G, H): one-year old juvenile. (A, C, E, G): fluorescent micrographs; (B, D, F, H): corresponding phase-contrast pictures. bst, branchiostegite; cu, cuticle; gf, gill filaments; hl, hemolymph lacuna; iep, inner-side limiting epithelium; lm, lamina; n, nucleus; oep, outer-side limiting epithelium; pc, pillar cells. On the fluorescent micrographs, arrows indicate Na-K-ATPase reactivity. Bars: 40 μm.

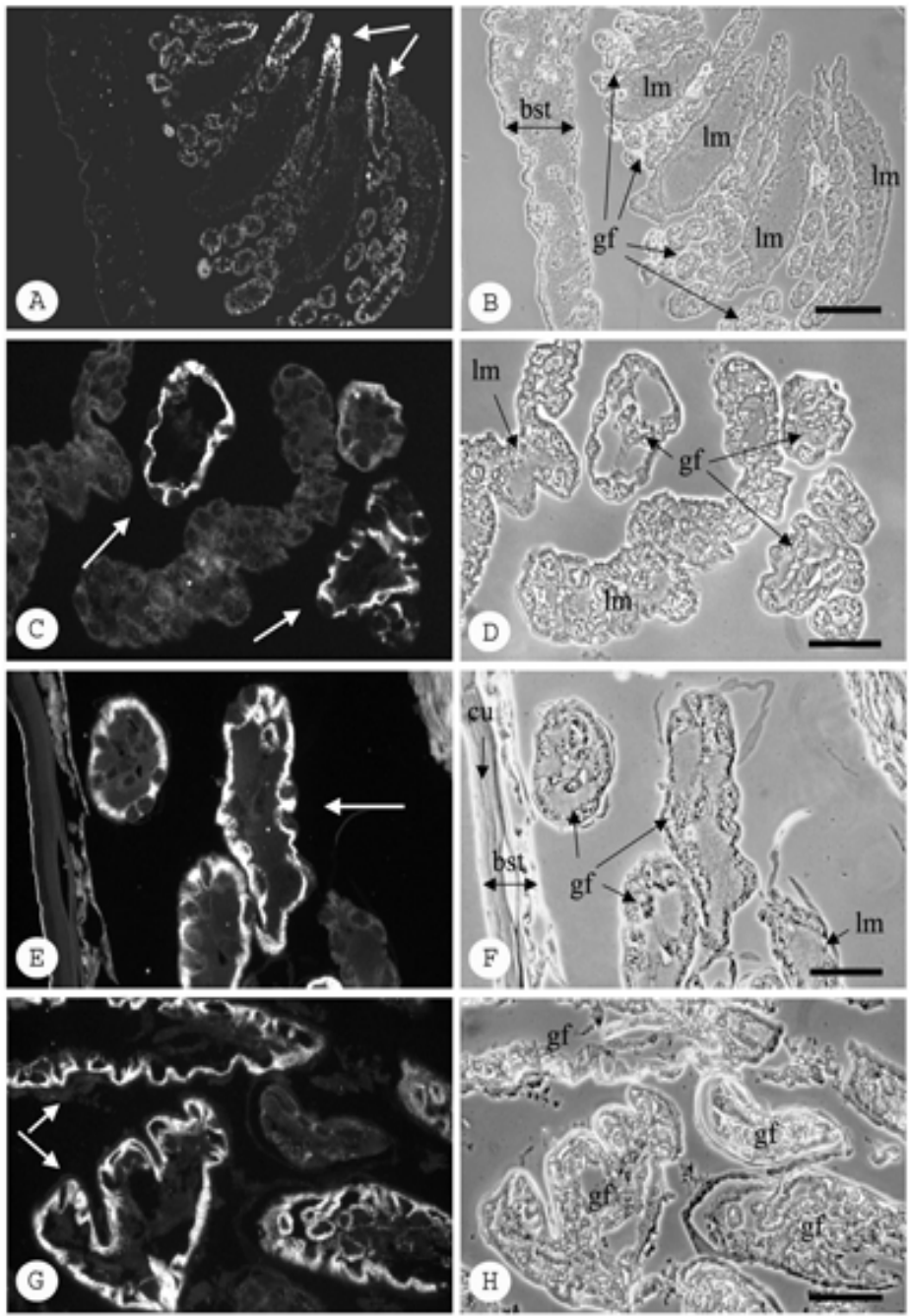